\documentclass[11pt,a4paper]{article}

\pdfoutput=1
\usepackage{amsmath,amssymb,authblk,multirow,collref}
\usepackage{graphicx}
\usepackage{pgfplots}
\usepackage{xparse}
\usepackage{subfigure}
\usepackage{float}

\usepackage{tikz}
\usetikzlibrary{positioning,arrows,patterns}
\usetikzlibrary{decorations.pathmorphing}
\usetikzlibrary{decorations.markings}
\usetikzlibrary{calc}
\tikzset{
    photon/.style={decorate, decoration={snake}, draw=black, thick},
    fermionnoarrow/.style={draw=black, postaction={decorate}, thick},
    scalar/.style={draw=black, postaction={decorate}, thick, dashed},
    fermion/.style={draw=black, postaction={decorate},decoration={markings,mark=at position .55 with {\arrow{>}}}, thick},
    gluon/.style={decorate, draw=black, decoration={coil,amplitude=4pt, segment length=5pt}, thick},
    vertex/.style={draw,shape=circle,fill=black,minimum size=3pt,inner sep=0pt} 
}

\graphicspath{{./}{./Figures/}}

\usepackage{geometry}
\newcommand{\lag}{\mathcal{L}}

\newcommand{\lhs}{\lambda_{HS}}

\newcommand{\vew}{v_{EW}}

\newcommand{\set}[1]{\mathbb{#1}}
\newcommand{\abs}[1]{\left\lvert #1 \right\rvert}

\newcommand{\ScaSinSADM}{1211.1014}

\newcommand{\modeqref}[1]{Eq.~\eqref{#1}}

\newcommand{\secref}[1]{Section~\ref{#1}}

\newcommand{\refcite}[1]{Ref.~\cite{#1}}
\newcommand{\refscite}[1]{Refs.~\cite{#1}}
\newcommand{\figref}[1]{Figure~\ref{#1}}

\usepackage[sort&compress,numbers,colon]{natbib}
\usepackage{hyperref}

\begin{document}

\title{The Galactic Center Excess from $\set{Z}_3$ Scalar Semi-Annihilations}
\author[1]{Yi Cai\thanks{\texttt{yi.cai@unimelb.edu.au}}}
\author[2]{Andrew Spray\thanks{\texttt{a.spray.work@gmail.com}}}
\affil[1]{ARC Centre of Excellence for Particle Physics at the Terascale, School of Physics, 
The University of Melbourne, Victoria 3010, Australia}
\affil[2]{Center for Theoretical Physics of the Universe, Institute for Basic Science (IBS), Daejeon, 34051, Korea}

\maketitle

\begin{abstract}
We investigate the possibility of the $\set{Z}_3$ scalar singlet model explaining the Fermi galactic centre excess.  We find a good fit to the measured spectral excess in the region where the dark matter mass is comparable to the Higgs and the Higgs portal coupling $\lambda_{HS}\sim 0.04$. This preferred region is consistent with constraints from vacuum stability and current dark matter experiments, and will be discovered or falsified soon by future dark matter direct detection experiment.   
\end{abstract}

\section{Introduction}

The nature of dark matter (DM) is one of the biggest questions in contemporary particle physics.  Despite intense efforts at many different experiments, no \emph{unambiguous} non-gravitational signal has been found.  However, current and near future experiments offer the prospect of decisively testing the weakly-interacting massive particle (WIMP) hypothesis.  This makes experimental excesses more theoretically attractive now than in the past.

One of the most interesting recent anomalies in dark matter searches is the $\gamma$-ray galactic centre excess (GCE) discovered in Fermi data after subtraction of backgrounds.  The original discovery~\cite{1010.2752} has been corroborated by several theoretical analyses~\cite{1012.5839,1110.0006,1207.6047,1402.6703,1404.3218,1409.0042,1411.2592}, and recently by the Fermi collaboration itself~\cite{1511.02938}.  The GCE has many of the features expected of DM annihilation into Standard Model (SM) states: the morphology in the sky matches what is expected from DM density distributions, and the required cross sections are very close to that of the canonical thermal WIMP.  The spectrum is easily fit by various SM final states; $b\bar{b}$ for DM masses of 30-60~GeV offers the best fit, but Higgs, gauge boson and top final states with larger DM masses are also acceptable~\cite{1411.2592,1411.4647,1503.08213} .  This has inspired considerable model building efforts to explain the observed signal~\cite{1206.5779,1401.6458,1403.5027,1404.0022,1404.1373,1404.4977,1404.6528,1405.0272,1405.4877,1406.3598,1407.0174,1409.1573,1410.4842,1501.00206,1501.02666,1501.03507,1502.05703,1503.01773,1504.03610,1510.00714,1510.07562}.  Alternative explanations have been advanced, including pulsars~\cite{1309.3428,1402.4090,1404.2318,1411.2980,1504.02477,1506.05104} or cosmic rays at the galactic centre~\cite{1405.7928,1410.7840,1506.05119}, but the efficacy of these explanations is contested~\cite{1305.0830,1402.6703,1407.5625,1509.02928}.  In any case, we find it interesting to consider a DM interpretation of this signal.

\begin{figure}
  \centering
  \begin{tikzpicture}[node distance=1cm and 1.75cm]
    \coordinate (v1);
    \coordinate[above left = of v1, label=above left:$\chi$] (i1);
    \coordinate[below left = of v1, label=below left:$\chi$] (i2);
    \coordinate[above right = of v1, label=above right:{$V$}] (o1);
    \coordinate[below right = of v1, label=below right:{$V$}] (o2);
    \draw[fermionnoarrow] (i1) -- (v1);
    \draw[fermionnoarrow] (v1) -- (i2);
    \draw[fermionnoarrow] (o2) -- (v1);
    \draw[fermionnoarrow] (v1) -- (o1);
    \draw[fill = white] (v1) circle (1);
    \fill[pattern = north west lines] (v1) circle (1);
  \end{tikzpicture}\quad
  \begin{tikzpicture}[node distance=1cm and 1.75cm]
    \coordinate (v1);
    \coordinate[above left = of v1, label=above left:$\chi_1$] (i1);
    \coordinate[below left = of v1, label=below left:$\chi_2$] (i2);
    \coordinate[above right = of v1, label=above right:{$\chi_3$}] (o1);
    \coordinate[below right = of v1, label=below right:{$V$}] (o2);
    \draw[fermionnoarrow] (i1) -- (v1);
    \draw[fermionnoarrow] (v1) -- (i2);
    \draw[fermionnoarrow] (o2) -- (v1);
    \draw[fermionnoarrow] (v1) -- (o1);
    \draw[fill = white] (v1) circle (1);
    \fill[pattern = north west lines] (v1) circle (1);
  \end{tikzpicture}
  \caption{DM annihilation (left) and SA (right), where $\chi$ ($V$) is a dark (visible) sector field.}\label{fig:SA}
\end{figure}
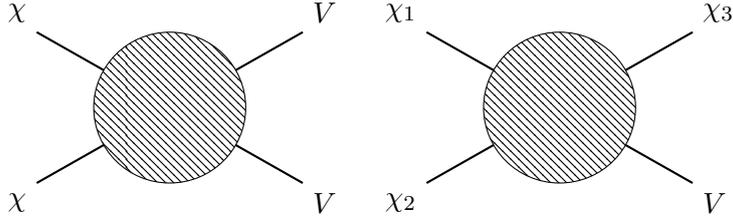

Semi-annihiatlon (SA) is a generic feature in DM phenomenology that occurs whenever DM is stabilized by a symmetry larger than $\set{Z}_2$.It specifically modifies the relic density and indirect detection signals, which makes it interesting for interpreting the GCE.  We compare SA and ordinary DM annihilation in \figref{fig:SA}.  SA is characterised by non-decay processes with an odd number of external dark-sector states.  In addition to enabling different final states and kinematics, SA is also irrelevant for collider and dark matter searches.  This can allow strong indirect signals while weakening other constraints.

In a previous work~\cite{1509.08481}, we showed that SA in a two-component DM model could explain the GCE through processes where a single Higgs is produced nearly at rest.\footnote{For other attempts to explain the GCE in SA models, see \refscite{1407.5492,1502.00508,1507.08295}.}  The subsequent decay of the Higgs to $b\bar{b}$ produces a spectrum very similar to annihilation of 60~GeV DM to the same final state.  The dark sector only coupled to the SM through the Higgs portal, and SA played two essential roles related to this interatction.  With SA significantly contributing to setting the thermal relic density, smaller Higgs portal couplings were allowed, thereby alleviating the stringent constraints from LUX.  Additionally, DM annihilating through the Higgs portal will always preferentially produce gauge bosons over Higgses, resulting in a poorer fit to the GCE $\gamma$-ray spectrum.  A large SA cross section substantially enhances the production rate of Higgses in our model. 

In this paper, we apply the same approach to the simplest SA model: the $\set{Z}_3$ scalar singlet model (Z3SSM)~\cite{\ScaSinSADM}.  This model extends the SM by a single DM particle only, coupled renormalisably to the visible sector through the Higgs portal.  It is also a minimal extension of the well-studied $\set{Z}_2$ scalar singlet model~\cite{Silveira:1985rk}.
The benefits that SA brought in our previous model apply also to the Z3SSM: weakening the direct detection bounds and enhancing the production rate of Higgses.  The Z3SSM is a simpler and more constrained model and it is of interest to see whether it also enjoys the privilege to explain the GCE. 

The outline of this paper is as follows.  We begin by reviewing the Z3SSM and vacuum stability constraints in \secref{sec:model}.  Our fit to the GCE is described in \secref{sec:gce}, and other relevant constraints are given in \secref{sec:cons}.  Finally we give our conclusions in \secref{sec:conc}.

\section{Model}\label{sec:model}

The Z3SSM is the simplest example of semi-annihilating dark matter.  The SM is extended by one new particle, a gauge singlet complex scalar $S$.  Its stability is ensured by a global $\set{Z}_3$ symmetry under which it is charged while all SM fields are neutral.  This can be the low-energy remnant of a spontaneously broken $U(1)$ gauge symmetry, provided that $S$ is the only light degree of freedom~\cite{1402.6449}.  The Lagrangian is
\begin{equation}
  \lag = \lag_{SM} + (\partial_\mu S)^\dagger (\partial^\mu S) - (M_S^2 - \frac{1}{2} \, \lhs \vew^2) S^\dagger S - \frac{\mu_3}{2} \bigl(S^3 + (S^\dagger)^3\bigr) - \lhs (H^\dagger H) (S^\dagger S) - \frac{1}{2} \lambda_{S} (S^\dagger S)^2 .
\end{equation}
There are four new parameters: the DM mass $M_S$, SA coupling $\mu_3$, Higgs portal coupling $\lhs$ and quartic $\lambda_S$.  Of these, $\mu_3$ may be taken real and positive without loss of generality while the quartic couplings are perturbative if $\abs{\lhs} < 4\pi$ and $\abs{\lambda_S} < \pi$~\cite{0909.0520}.

The vacuum stability of this model was studied in \refcite{\ScaSinSADM}.  There are four inequivalent vacua, depending on which of $H$ and $\phi$ acquire VEVs.  The desired solution has $\langle \phi \rangle = 0$ to preserve the global $\set{Z}_3$, and $\langle H \rangle \neq 0$ for EWSB.  Demanding that this is the global minimum at the weak scale sets the bound 
\begin{equation}
  \mu_3 \lesssim 2\sqrt{\lambda_S} M_S \leq 2 \sqrt{\pi}\, M_S \,.
\label{eq:mu3cons}\end{equation}
The second inequality results from imposing the perturbativity constraint.  This is the most robust consistency bound on the SA coupling.\footnote{Allowing this vacuum to be metastable with a sufficiently long lifetime weakens the bound slightly, but does not significantly affect our analysis.}  Stronger bounds exist when we consider the effects of the renormalisation group.  For the values of the Higgs-portal couplings of interest, $\lhs \approx 0.05$, large values of $\lambda_S$ destabilize the vacuum as we run the couplings to a high scale $\Lambda$.  Requiring that $\Lambda \lesssim 10^9$~GeV, \emph{i.e.} the same scale as the instability in the SM,  gives the stronger constraint
\begin{equation}
  \lambda_S \lesssim 0.5\,, \qquad \mu_3 \lesssim \sqrt{2} \, M_S \,.
\label{eq:mu3consstronger}
\end{equation}
Aside from these vacuum stability constraints, the DM quartic $\lambda_S$ has no further effect on the phenomenology, leaving a 3-dimensional parameter space.

\section{Galactic Center Excess}\label{sec:gce}

The Planck satellite measured the dark matter density $\Omega_{DM} h^2 \in [0.1126, 0.1246]$ at 3$\sigma$~\cite{1502.01582}.  We use micrOMEGAs~4.0~\cite{1407.6129} to compute the relic density including the effect of SA, and demand that $S$ saturate this result.  This fixes $\mu_3$ as a function of $(M_S, \lhs)$.  In this model, SA decreases the relic abundance, so there is an effective upper bound on $\lhs$ as a function of $M_S$ given when the correct relic abundance is obtained for $\mu_3 = 0$.  This is the upper grey region in \figref{fig:j4}, for $\lhs \gtrsim 0.06$.  We also show a lower grey region where we need $\mu_3 > 4000 $ GeV to produce the Planck results, and a gold line showing the consistency bound of \modeqref{eq:mu3consstronger}.

The differential photon flux from DM annihilation and SA is
\begin{equation}
  \frac{d^2\Phi}{d\Omega dE_\gamma} = \frac{1}{8\pi m_S^2} \sum_i \frac{dN_i}{dE_\gamma} \, \langle\sigma_i v\rangle \int_{l.o.s.} \rho^2 dl \,,
\label{eq:genID}\end{equation}
where $\rho$ is the DM density and the sum $i$ runs over different annihilation and SA channels.  The astrophysical dependence of the flux factors out, and is conventionally expressed in terms of 
\begin{equation}
  \bar{J} (\Delta \Omega) = \frac{1}{\Delta \Omega} \int_{\Delta\Omega} d\Omega \int_{l.o.s.} \rho^2 dl \,.
\end{equation}
We follow \refcite{1409.0042} and parameterise $\bar{J}$ as
\begin{equation}
  \bar{J} = \mathcal{J} \, \bar{J}_{can} \,,
\end{equation}
where $\bar{J}_{can} = 1.58\times 10^{24}$~GeV$^2$/cm$^5$.   $\mathcal{J} \in [0.14, 4.0]$ represents our uncertainty in the DM density distribution.  

The sum in \modeqref{eq:genID} runs over all annihilation processes $S^\dagger S \to SM$, plus the SA process $SS \to S^\dagger h$.  We use the PPPC~4~DM~ID~\cite{1012.4515} expressions $f_i (m_{DM}, E_\gamma)$ for the energy distributions.  For the SA channel, we use the PPPC results for the $hh$ final state, multiplied by one-half as only a single Higgs is produced.  Additionally the Higgs is produced with an energy
\begin{equation}
  E_h (m_S) = \frac{3 m_S^2 + m_h^2}{4 m_S} \,,
\label{eq:Eh}\end{equation}
so we use the PPPC distributions for $m_{DM} = E_h$.  We may then write the total photon flux as
\begin{equation}
  \frac{d\Phi}{dE_\gamma} (E_\gamma) = \frac{\mathcal{J} \, \bar{J}_{can} \Delta \Omega}{8\pi m_S^2} \Biggl( \sum_{i,ann} \langle\sigma_i v\rangle \, f_i (m_S, E_\gamma) + \frac{1}{2} \langle\sigma_{SA} v\rangle \, f_{hh} (E_h, E_\gamma ) \Biggr) \,.
\end{equation}

\begin{figure}
\centering
\includegraphics[width=0.6\textwidth]{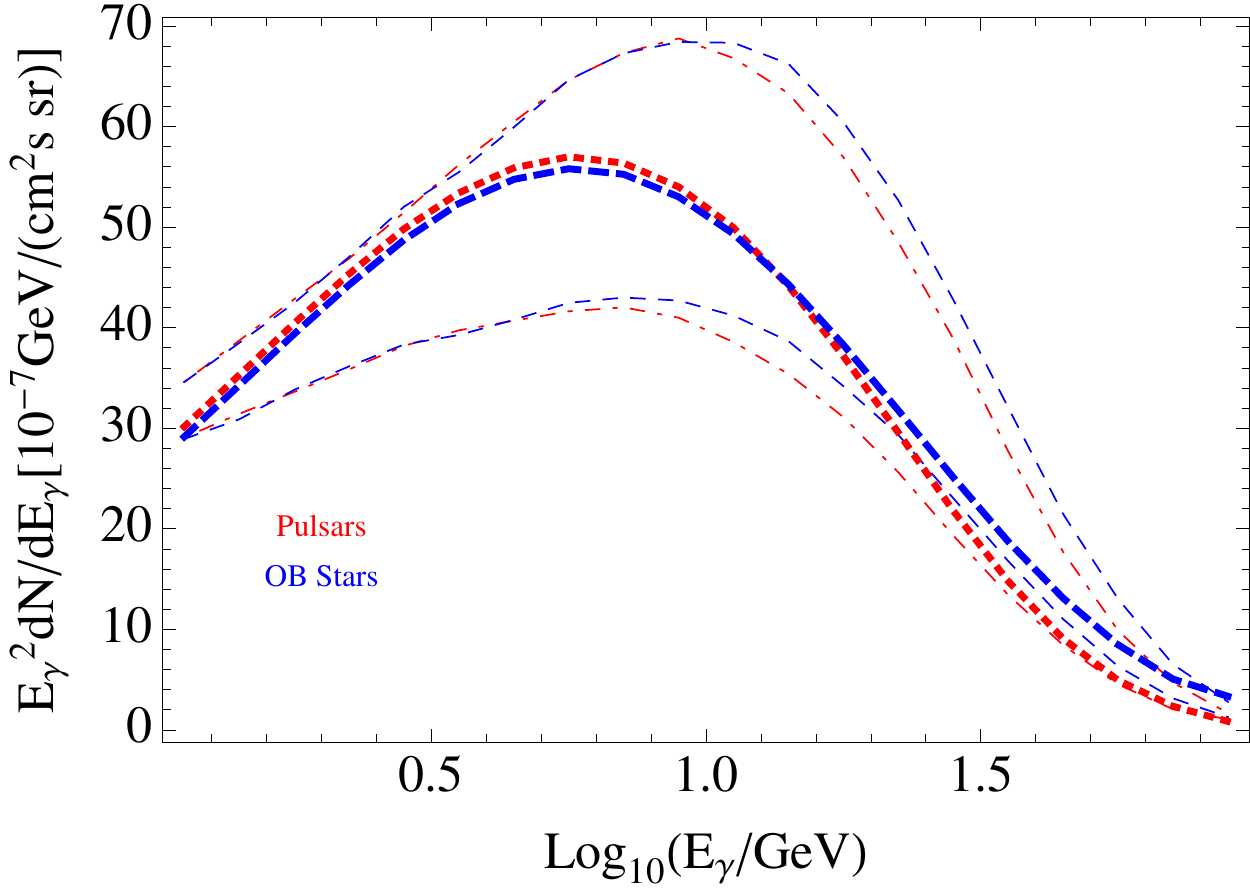}
\caption{The photon flux for the best-fit point in the Pulsar (OB Stars) intensity-scaled scenario is shown by the red dotted (blue) line.  The Fermi residual spectral bands lie between the fainter red dot-dashed (blue dashed) lines.}
\label{fig:bestfit}
\end{figure}

We perform a $\chi^2$ analysis on the computed photon spectum in 20 bins with the observed Fermi spectrum~\cite{1511.02938} assuming symmetric uncertainty distribution.  The quality of the fit improves with increasing $\mathcal{J}$, especially once all constraints are considered. The best fits are found with spectra modelled with an exponential cut-off power law where intensity-scaled Pulsars and OB-stars models account for interstellar gamma ray emission.\footnote{The Fermi collaboration~\cite{1511.02938} considered intensity-scaled and index-scaled models for this background; the former varied the normalisation only, while the latter also varied the spectral index.}  The best fit point for the pulsar (OB stars) intensity-scaled scenario with $\mathcal{J} = 4.0$ is $m_S =125 (161)$ GeV, $\lhs =0.0467 (0.0548)$, $\mu_3 = 52.7 (64.5)$~GeV and $\chi^2 = 2.4 (2.1)$.  We show the flux at these points together with the GCE spectra in \figref{fig:bestfit}, and the 1, 2 and 3$\sigma$ contours in \figref{fig:j4}.  

Only one of the other interstellar emission models considered by the Fermi collaboration can be reasonably fit with the Z3SSM: the index-scaled Pulsar scenario where the excess is modelled with a power law without a cut-off.  The fit here is worse, but the best fit point has $\chi^2 = 16.1/20$~d.o.f. (for $\mathcal{J} = 4.0$ and $m_S =132$ GeV, $\lhs =0.0335$, and $\mu_3 = 86.7$~GeV).  The phenomenology of this case is otherwise very similar to the two shown in \figref{fig:j4}.  The fit to other interstellar emission models considered by the Fermi collaboration all have $\chi^2/$d.o.f.$\,\gtrsim 4$. Spectra modelled using a power law (without a cut-off) are too hard for the Z3SSM, while those based on index-scaled interstellar emission models are too soft.

The best fit point is characterised by a strong annihilation signal and a relatively small SA cross section.  The contribution of the different final states $WW:ZZ:hh:S^\dagger h \approx 69\%:29\%:1.2\%:0.43\% (45\%:20\%:27\%:8.0\%)$ for Pulsars (OB stars).  This arises as the SA cross section is bound by the perturbativity constraints on $\mu_3$.  However, SA still plays an important role in setting the relic density for smaller $\lhs$, and thus avoiding direct detection constraints. 

\section{Additional Constraints}\label{sec:cons}

An unavoidable constraint for any potential explanation of the GCE comes from Fermi observations of dwarf spheroidal galaxies (dSphs)~\cite{1410.2242,1510.06424}.  These are DM-dominated objects offering low backgrounds and reasonably well-understood density distributions.  The same experiment that observed the GCE has seen no corresponding excess from these sources.  

\begin{figure}
\centering
\includegraphics{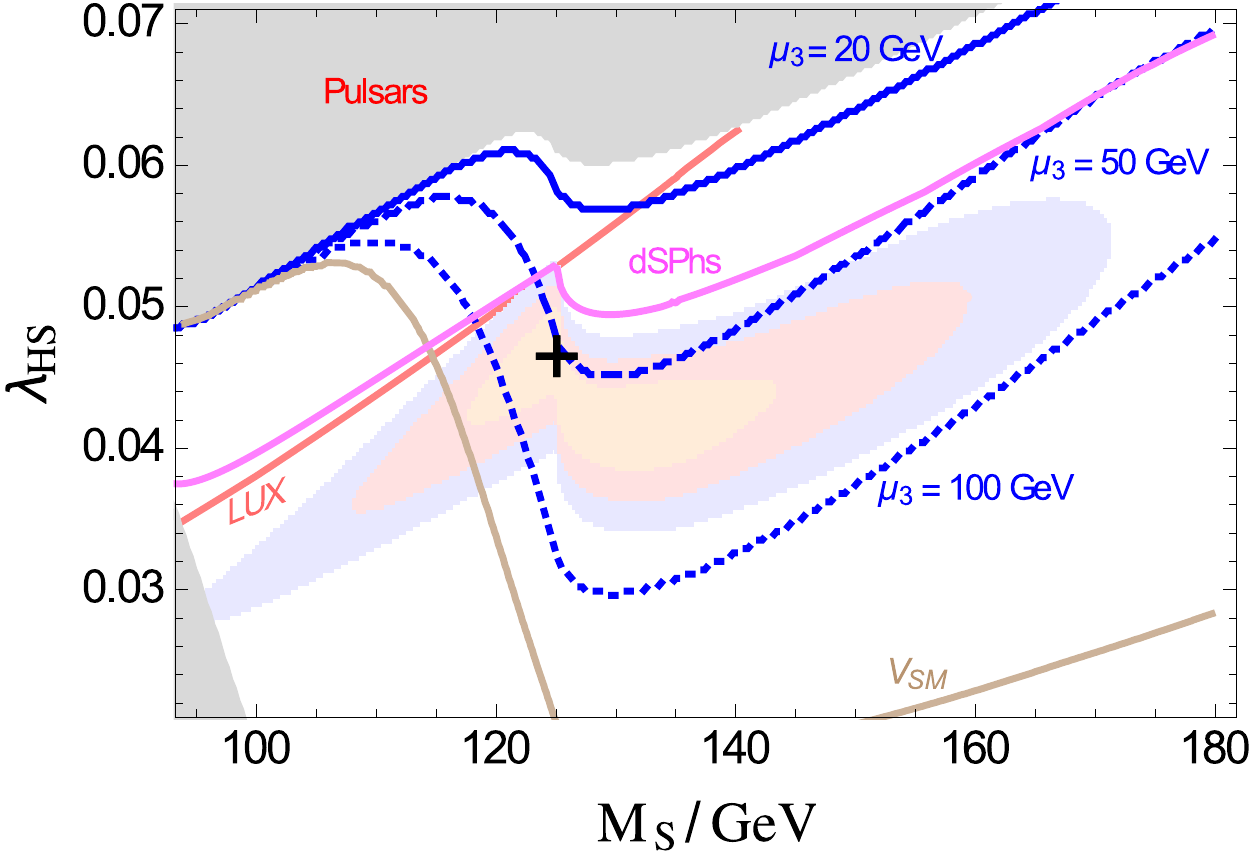}\\
\includegraphics{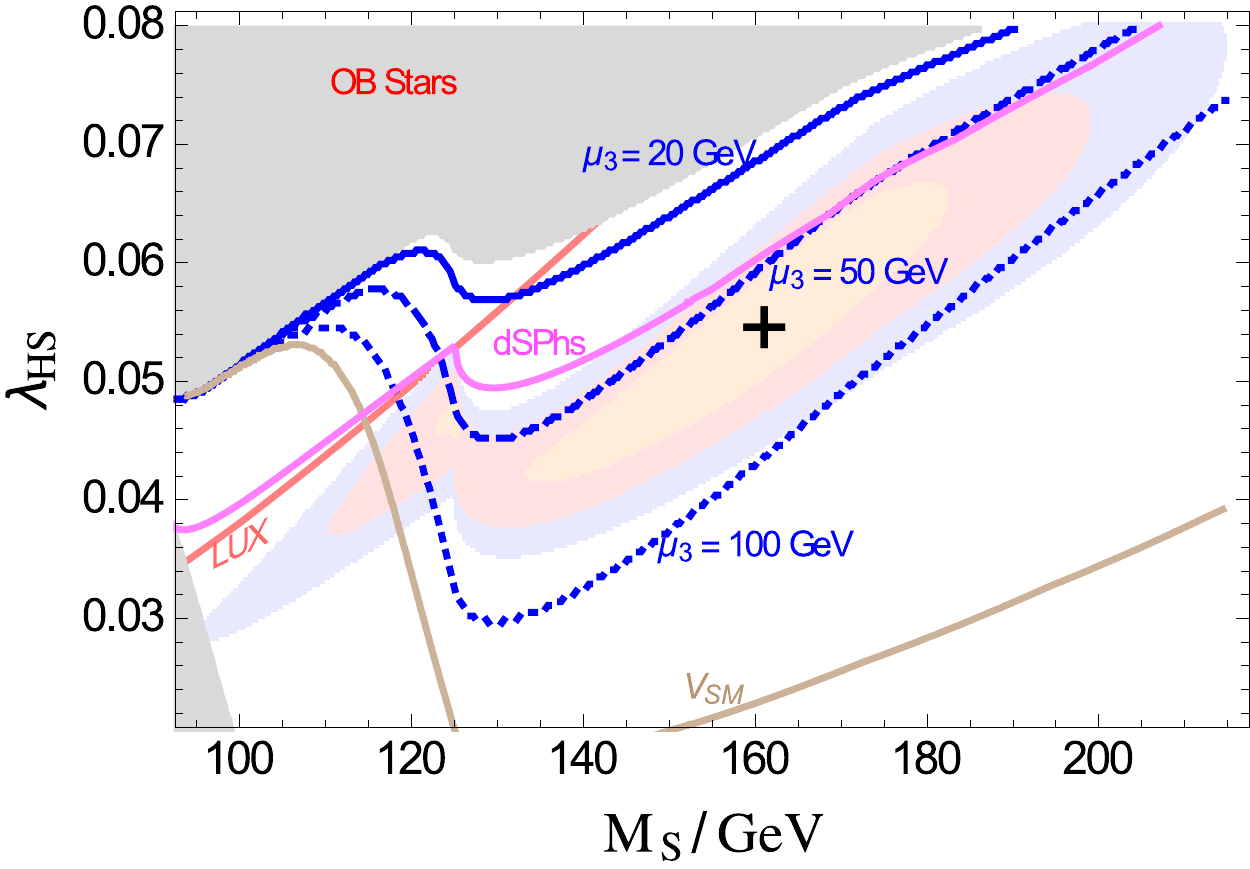}
\caption{Preferred region for the GCE with experimental and theoretical constraints in the $M_S$--$\lhs$ plane for two interstellar emission models: Pulsars intensity-scaled (upper panel) and OB Stars intensity-scaled (lower panel).  In the grey regions, no value of $\mu_3 \in [0, 4000]$ GeV  could reproduce the observed relic density.  Contours of $\mu_3$ are in blue, with the gold line labelled $V_{SM}$ denoting the stability bound of \modeqref{eq:mu3consstronger}.  The black cross marks the GCE best-fit point, and the yellow, red and blue shaded ellipses the 1, 2 and 3-$\sigma$ contours.  Bounds from LUX and from Fermi observations of dSphs, as discussed in \secref{sec:cons}, are as labelled.}
\label{fig:j4}
\end{figure}

The most recent limits from Fermi~\cite{1503.02641} are given in terms of cross sections to various SM final states.  Our model (semi-)annihilates to several different final states, so to apply these limits we make two simplifications.  We assume that the differences between the spectra for $WW$, $ZZ$ and $h(h)$ are sufficiently small that they do not significantly effect the constraints.  Further, since the SA channel is sub-dominant and the GCE preferred region is for $m_S \approx m_h$, we neglect the effects of the different Higgs boost of \modeqref{eq:Eh}.  This gives us the bound
\begin{equation}
  \sum_{i,ann} \langle\sigma_i v\rangle + \frac{1}{2} \langle\sigma_{SA} v\rangle < \sigma v^{dSphs,WW} .
\end{equation}
We show this in \figref{fig:j4} in pink.  We see that for $\mathcal{J} = 4$, the best-fit region is not excluded.  However, smaller values $\mathcal{J} \lesssim 3$ start to be in tension.  This is in line with other studies that have found that the GCE and dSph bounds are consistent only for an enhanced signal from the galactic center~\cite{1510.06424}.

Limits from collider searches and direct detection are independent of SA (and $\mu_3$), and similar to bounds on for a scalar singlet $\set{Z}_2$ model.  Collider searches in monojets and jets + MET set no current bounds in the mass range of interest, and are not expected to be constraining in the near future~\cite{1509.08481,1407.6882,1412.0258}.  

Direct detection bounds in contrast are very relevant.    The spin-independent scattering cross section is
\begin{equation}
  \sigma_{SI} (S N \to S N) = \frac{\lhs^2 f_N^2}{4\pi} \, \frac{m_N^4}{m_H^4 (m_N + M_S)^2} \,,
\end{equation}
where $f_N$ is the Higgs-nucleon coupling
\begin{equation}
  f_N = \sum_q f_q = \sum_q \frac{m_q}{m_N} \, \langle N \lvert \bar{q} q \rvert N \rangle \,,
\end{equation}
and $m_N = 0.946$~GeV.  We follow \refcite{Cline:2013gha} and take $f_N = 0.345$ in placing our limits.  See \refscite{Cline:2013gha,Alarcon:2011zs,Alarcon:2012kn,Junnarkar:2013ac,Ellis:2008hf,Akrami:2010dn,Bertone:2011nj,Crivellin:2013ipa,Hoferichter:2015dsa,Alarcon:2012nr} for more details.  The strongest spin-independent scattering limits for DM masses $\gtrsim 5$~GeV are from the preliminary run at LUX~\cite{Akerib:2013tjd}.  For $m_S = 140$~GeV, LUX excludes $\sigma_n > 1.7 \times 10^{-45}$~cm$^2$.  We plot the resultant limits in \figref{fig:j4} in orange.  For $m_S < m_h$, these are the most stringent limits on $\lhs$. 

\section{Conclusions}\label{sec:conc}

In this paper, we studied the simplest example of semi-annihilating DM, the Z3SSM, in the context of the Fermi GCE.  The Z3SSM has only three parameters relevant to DM phenomenology; we reduce this to a two-dimensional parameter space by demanding the relic density reproduce observations.  Our main result is given in \figref{fig:j4}, which shows the best-fit regions as well as constraints from direct detection, dwarf spheroidals and vacuum stability.

We find that, assuming a substructure enhancement in the signal from the galactic centre, this model \emph{can} explain the GCE while remaining consistent with all current constraints.  The agreement with the measured spectral excess is quite good for a 20-bin analysis with $\chi^2 = 2.1$ at the best fit point and spectrum.  SA is a subdominant contribution to the $\gamma$-ray flux, but plays an essential role in obtaining the correct relic density and easing bounds from LUX and from dwarf spheroidals.  

Finally, we note that the region where this model explains the GCE only marginally evades the current bounds.  Moderate improvements in either type of limit should exclude or, more optimistically, discover the region of parameter space discussed here.  In particular, the full LUX results are expected to improve the direct detection bounds by a factor of $\sim 5$ in the relevant mass range~\cite{Szydagis:2014xog}. Thus we expect this explanation of the GCE to be proved or falfisied soon.

\section*{Acknowlegements}
AS would like to thank the Instituto de Fisica Teorica (IFT UAM-CSIC) in Madrid for its support via the Centro de Excelencia Severo Ochoa Program under Grant SEV-2012-0249, during the IBS-Multidark Joint Workshop on Dark Matter, where part of this work was completed. YC was supported by the Australian Research Council.  This work was supported by IBS under the project code, IBS-R018-D1.

\bibliography{Z3min_GCE-v2}{}
\bibliographystyle{JHEP}

\end{document}